\documentclass[notitlepage,twocolumn,prl,tightenlines,nofootinbib,superscriptaddress]{revtex4}

\pdfoutput=1

\usepackage{amsmath}
\usepackage{amssymb,amsfonts}
\usepackage{graphicx}
\usepackage{bm}
\usepackage{subfigure}
\graphicspath{{figs/}}

\newcommand{\ie}{\textit{i.e.}}

\newcommand{\mathnotation}[2]{\newcommand{#1}{\ensuremath{#2}}}

\newcommand{\Order}[1]{\mathrm{O}(#1)}      

\newcommand{\<}{\langle}
\renewcommand{\>}{\rangle}
\renewcommand{\time}{t}
\mathnotation{\ee}{\mathrm{e}}
\mathnotation{\xv}{\bm{x}}                  
\mathnotation{\uv}{\bm{u}}                  
\mathnotation{\ruv}{\hat{\bm{r}}}           
\mathnotation{\Uc}{U}                       
\mathnotation{\Lsc}{L}                      
\mathnotation{\lsc}{\ell}                   
\mathnotation{\Nswim}{N}                    
\mathnotation{\nd}{n}                       
\mathnotation{\nV}{c}                       
\mathnotation{\ai}{a}                       
\mathnotation{\Nreal}{M}                    
\mathnotation{\Square}{\cal S}              
\mathnotation{\sdim}{d}                     
\mathnotation{\pal}{\lambda}                

\mathnotation{\W}{\;\mathrm{W}}
\mathnotation{\cm}{\;\mathrm{cm}}
\mathnotation{\meter}{\;\mathrm{m}}
\mathnotation{\kg}{\;\mathrm{kg}}

\begin{document}

\title{Stirring by swimming bodies}

\author{Jean-Luc Thiffeault}
\affiliation{Department of Mathematics, University of Wisconsin --
  Madison, 480 Lincoln Dr., Madison, WI 53706, USA}
\affiliation{Institute for Mathematics and Applications, University of
  Minnesota -- Twin Cities, 207 Church Street S.E., Minneapolis, MN
  55455, USA}
\email{jeanluc@math.wisc.edu}
\author{Stephen Childress}
\affiliation{Courant Institute of Mathematical Sciences, New York
  University, 251 Mercer Street, New York, NY 10012, USA}



\begin{abstract}
  We consider the stirring of an inviscid fluid caused by the
  locomotion of bodies through it.  The swimmers are approximated by
  non-interacting cylinders or spheres moving steadily along straight
  lines.  We find the displacement of fluid particles caused by the
  nearby passage of a swimmer as a function of an impact parameter. We
  use this to compute the effective diffusion coefficient from the
  random walk of a fluid particle under the influence of a
  distribution of swimming bodies.  We compare with the results of
  simulations.  For typical sizes, densities and swimming velocities
  of schools of krill, the effective diffusivity in this
  model is five times the thermal diffusivity.  However, we estimate
  that viscosity increases this value by two orders of magnitude.
\end{abstract}

\maketitle 

\citet{Munk1966} was the first to ask whether biology has an important
impact on mixing in the oceans.  Since mixing affects the global
circulation and stratification of the oceans, it is of great interest
to physical oceanographers to settle this question.  \citet{Dewar2006}
have proposed that the mechanical energy delivered by the swimming
motions of the marine biosphere could amount to almost $10^{12}\W$, a
figure comparable to the energy delivered by the winds and tides.
This suggests a biological origin of about $33\%$ of the mixing in the
oceans, an enormous figure.  By assuming that this energy is delivered
to the top three kilometers of the oceans, they estimate an effective
diffusivity produced by swimmers to be approximately $.2 \cm^2/\sec$,
or about 100 times the molecular value for heat.  \citet{Kunze2006}
have measured elevated levels of ocean turbulence due to swimming
krill, though some have questioned whether this turbulence can
efficiently overturn a stratified
medium~\cite{Visser2007,Gregg2009}. \citet{Katija2009} suggest that
the displacement of fluid particles by swimming bodies, which viscous
effects can lengthen, is more relevant to stirring than the scale of
turbulence they produce.  This is the viewpoint adopted in this
Letter.

\citet{Huntley2004} considered the energy produced by~$11$
representative species of schooling animals, from krill to whales, and
found that regardless of size the energy input to the ocean per unit
mass from the swimming of these animals was roughly a constant of
order $10^{-5} \W/\kg$.  If the average the biomass of the oceans has
a volume fraction $\nV$ over a volume $V$ in cubic meters, we can thus
(assuming biological materials have the density of water) arrive at a
total energy input of $10^{-2} \nV V\W$. The area of the oceans is about
$3.6\times 10^{14}\meter^2$, and if we assume the biomass is evenly
distributed to a depth $D$ meters, we get an input of $3.6 D\nV\times
10^{12}$ W. This suggests that for a typical depth of one kilometer
the volume density of biomass should be of order
$10^{-3}$--$10^{-4}$, and thus that the organisms form a dilute
suspension, even if their distribution is patchy and organized into
schools.

The object of the present paper is to determine the effective
diffusivity and the statistics of the concentration field of a passive
scalar that can result from the fluid motions caused by such dilute
arrangements of swimming animals.  The scalar could be for example
heat, salt, or nutrients, and is assumed to have negligible feedback
on the flow (at least at small scales).  The animals considered by
\citet{Dewar2006} and \citet{Huntley2004} have large Reynolds number,
typically $10^2$--$10^7$.  Our focus is therefore distinct from the
mixing that can occur from dense, tightly interacting suspensions of
Stokesian swimmers, as in \cite{Pedley1992, Wu2000,
  HernandezOrtiz2006, Saintillian2007, Underhill2008, Leptos2009},
though it shares many common features such as linearity in the density
of swimmers.  Real swimmers present a wide variety of motions and
wakes, but these have generic forms in the far
field~\cite{Lighthill1991}.  For example, the far potential field of a
neutrally buoyant fish in steady unaccelerated swimming decays at
least like a quadrapole.  Following \cite{Katija2009}, we will model
the swimmers by identical cylinders or spheres moving in potential
flow, but these are merely stand-in examples.  We emphasize that
though the motivating application comes from oceanography, the simple
model we introduce can be applied to a range of systems, such as the
mixing caused by vortices or suspensions of solid particles.

\textsl{Dilute suspensions of swimmers.}  As an extremely simple model
of stirring by swimmers, we consider a fluid particle, called the target
particle, which is influenced by the occasional passing of swimming
bodies. We assume that every swimmer moves at a fixed speed $\Uc$,
over distances $\Order{\Lsc}$ large compared to the ``range of
influence.''  This range will typically be a few body lengths normal
to the swimming path, and represents the distance from a target
particle where interaction with the swimmer becomes significant.  We
assume that encounters of our target particle with swimmers are
occasional, in that at each encounter the particle moves by a distance
$\Delta(\ai)$, where the impact parameter $\ai$ is the perpendicular
distance of the initially unperturbed particle from the extended line
of motion of the approaching swimmer. Since we are assuming that
$\Lsc$ is large compared to the distance of significant interaction,
we may assume that by an ``encounter'' we mean that the $\Delta(\ai)$
may be computed from motion of the swimmer along a doubly-infinite
line. Each encounter, the $k$th say, moves the particle a distance
$\Delta(\ai_k)$, in the direction of the unit vector~$\ruv_k$. Thus
after $\Nreal$ encounters the position of a target particle initially
at~$\xv_0$ is $\xv_\Nreal$ given by
\begin{equation}
  \xv_\Nreal = \xv_0 + \sum_{k=1}^{\Nreal} \Delta(\ai_k) \ruv_k\,.
  \label{eq:xvdot}
\end{equation}
Assuming infrequent encounters, we regard the $\ai_k, \ruv_k$ as
independent and identically distributed random variables.  We then
follow Einstein's derivation~\cite{Einstein1905} for computing the
displacement,
\begin{equation}
  \<|\xv|^2\> = \Nreal \<\Delta^2(\ai)\>
  + \Bigl\<\sum_{j\neq k}
  \Delta(\ai_j)\Delta(\ai_k)\,\ruv_j\cdot\ruv_k\Bigr\>,
  \label{eq:squaredisp}
\end{equation}
where we assume spatial homogeneity to set~$\xv_0=0$, and the angle
brackets denote ensemble averaging.  The $\ruv_k$ are isotropically
distributed, so the second term above vanishes.

To evaluate $\Nreal \<\Delta^2(\ai)\>$, consider a swimmer at a large
\begin{figure}
  \begin{center}
    \includegraphics[width=.65\columnwidth]{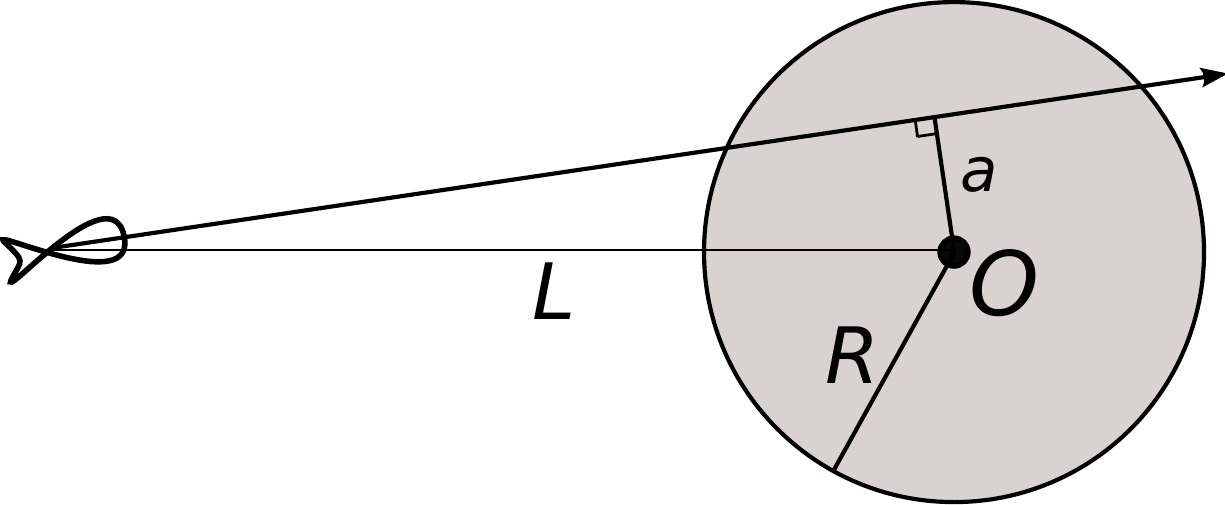}
  \end{center}
  \caption{A swimmer coming a distance~$\ai$ from a particle at~$O$.}
  \label{fig:hitting}
\end{figure}
distance~$\Lsc$ from the target particle at the origin~$O$ (see
Fig.~\ref{fig:hitting}).  Imagine a ``target disk'' (or sphere in 3D)
of radius~$R$, with~$\lsc \ll R \ll \Lsc$, where~$\lsc$ is a typical
length scale for the swimmer.  The fraction of swimmers that will hit
the target disk (sphere) is~$2R/2\pi\Lsc$ ($\pi R^2/4\pi\Lsc^2$ in
3D).  We can use the number density and the volume of the `shell' at a
distance~$\Lsc$ to find that the number of swimmers hitting the target
from a distance~$\Lsc$ is~$2R\nd\, d\Lsc$ ($\pi R^2\nd\, d\Lsc$ in
3D), where~$\nd$ is the number density of swimmers.  We now integrate
from~$0$ to~$\Uc t$, since swimmers further than~$\Uc t$ cannot hit
the target, to find the number of swimmers~$\Nreal$ that will hit the
interaction disk (sphere) in time~$t$: we find~$\Nreal = 2R\Uc\nd t$
($\Nreal = \pi R^2\Uc\nd t$ in 3D).  The expression for the squared
displacement~\eqref{eq:squaredisp} is now
\begin{equation*}
  \<|\xv|^2\> = \Nreal\int_0^R \rho(\ai)\Delta^2(\ai)d\ai
\end{equation*}
since the largest value of the impact parameter is~$R$.
Here~$\rho(\ai)$ is the probability distribution of impact parameters;
since the swimmers are assumed to arrive from far away, this
is~$\rho(\ai) = 1/R$ ($\rho(\ai) = 2\pi\ai/\pi R^2$ in 3D).  Combining
these results and taking~$R\rightarrow\infty$, we find
\begin{equation}
  \<|\xv|^2\>= \begin{cases}
    2\Uc\nd\, t \int_0^\infty \Delta^2(\ai)d\ai & ({\rm 2D});\\
    2\pi \Uc\nd\, t \int_0^\infty \ai\,\Delta^2(\ai)d\ai & ({\rm 3D}),
  \end{cases}
  \label{eq:squaredisp2D3D}
\end{equation}
assuming that the integrals converge.  Since the effective
diffusivity~$\kappa$ is defined by~$\<|\xv|^2\> = 2\sdim\kappa t$,
with~$\sdim$ the spatial dimension, Eq.~\eqref{eq:squaredisp2D3D} can
be used to determine~$\kappa$.

\textsl{Displacement due to a moving body}.
To find the displacement~$\Delta(\ai)$ of a target particle due to a
swimming body coming from infinitely far away and swimming in the~$z$
direction, we need to integrate the equations for the position of a
target particle at~$\xv(\time)$,
\begin{equation}
  \dot \xv = \uv(\xv,\time)\,,\qquad \xv(-\infty) = (\ai,0,-\infty)
  \label{eq:dispODE}
\end{equation}
for~$t$ ranging from $-\infty$ to $\infty$.  The impact parameter,
$\ai$, appears as the initial vertical position of the target
particle.  This classical `drift' problem has been treated in great
detail by many authors; see for example~\cite{Maxwell1869, Darwin1953,
  Eames1994, Eames1999}.

\begin{figure}
  \begin{center}
    \includegraphics[width=.8\columnwidth]{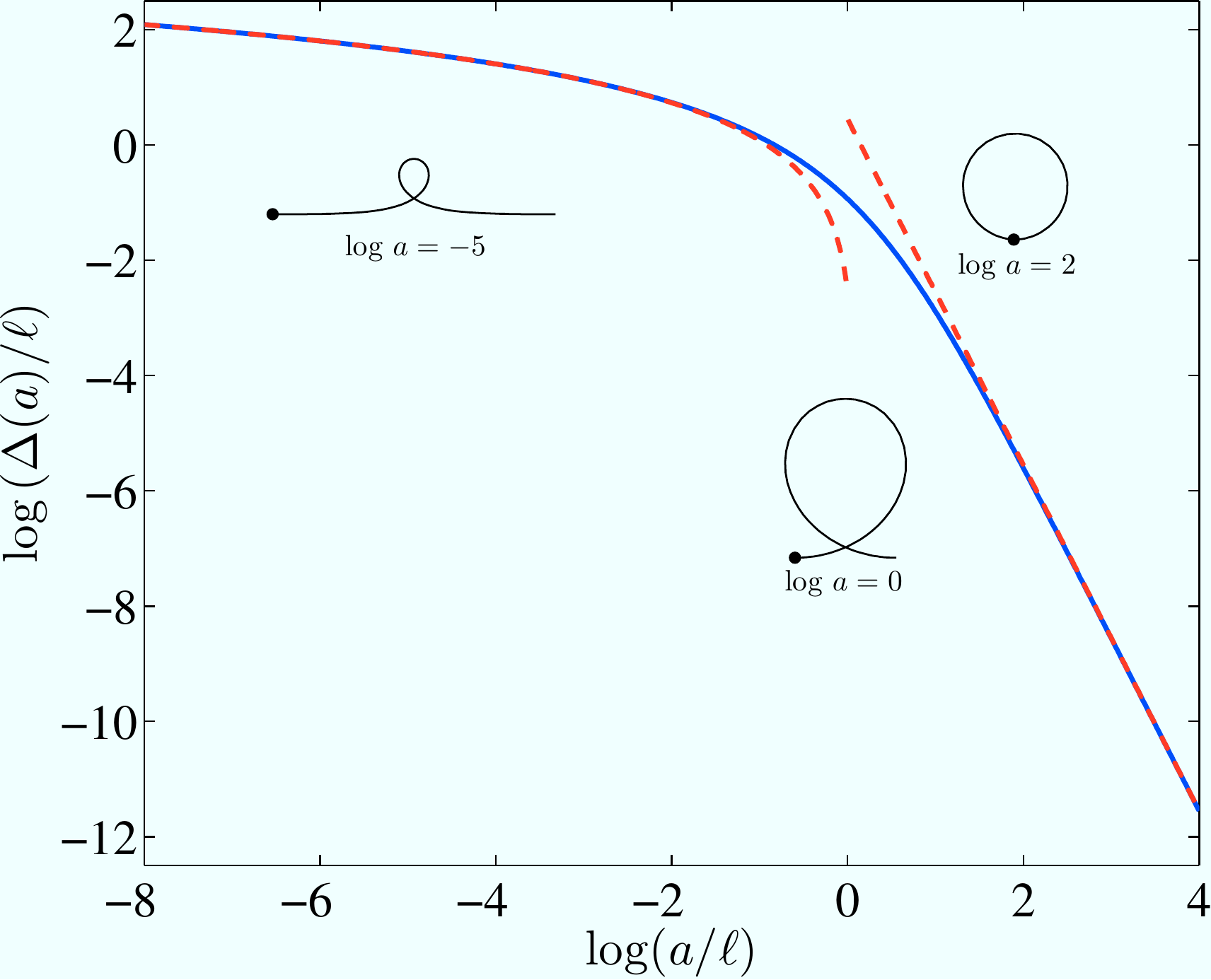}
  \end{center}
  \caption{Displacement~$\Delta(\ai)$ due to a cylinder, as a function
    of the impact parameter~$\ai$.  The dashed lines are the
    asymptotic forms for small and large~$\ai$ of
    Eq.~\eqref{eq:Deltacylasym}.  A few typical trajectories are
    overlaid: for large~$\ai$ the trajectory is almost circular, but
    is still ribbon-shaped.}
  \label{fig:Deltacyl}
\end{figure}

As a working example of the displacement due to a moving body, we
shall first treat the cylinder.  We consider the displacement in
two-dimensional flow due to the passage of a cylinder of radius~$\lsc$
moving at speed~$\Uc$ in potential flow.  Equation~\eqref{eq:dispODE}
can be integrated numerically for a given~$\ai$ to
obtain~$\Delta(\ai)$, plotted in Fig.~\ref{fig:Deltacyl}.  The
characteristic `ribbon' shape is evident.
Also shown are the asymptotic forms~\cite{Darwin1953}
\begin{equation}
  \Delta(\ai)/\ell \sim \begin{cases}
    \log(\ell/\ai) + .0794, & \quad \ai \ll \ell;\\
    \pi(\ell/\ai)^3/2, & \quad \ai \gg \ell.
  \end{cases}
  \label{eq:Deltacylasym}
\end{equation}
There is an integrable logarithmic singularity for small~$\ai$.  For
large~$\ai$, the trajectories are almost circles.  By combining these
asymptotic limits with numerical integration, we find the integral
from Eq.~\eqref{eq:squaredisp2D3D} is $\int_0^{\infty}
\Delta^2(\ai)d\ai \simeq 2.37\ell^3$.  But note that $\int_0^\ell
\Delta^2(\ai)d\ai \simeq 2.31\ell^3$: the integral is completely
dominated by ``head-on'' collisions ($97\%$ of the integral).

For a sphere in three-dimensional potential flow, the
displacement~$\Delta(\ai)$ appears superficially much as for the
cylinder in Fig.~\ref{fig:Deltacyl}, but drops off more rapidly for
large~$\ai$:
\begin{equation}
  \Delta(\ai)/\ell \sim \begin{cases}
    \frac43 \log(\ell/\ai) - .582, & \quad \ai \ll \ell;\\
    9\pi(\ell/\ai)^5/64, & \quad \ai \gg \ell.
  \end{cases}
  \label{eq:Deltasphasym}
\end{equation}
As for the cylinder, there is an integrable logarithmic singularity
for small~$\ai$, and the integral in Eq.~\eqref{eq:squaredisp2D3D} is
$\int_0^{\infty} \ai\Delta^2(\ai)d\ai \simeq .254\ell^4$.  The
small~$\ai$ logarithmic singularity is mollified by the extra factor
of~$\ai$ in the integral, but still~$\int_0^\ell \ai\Delta^2(\ai)d\ai
\simeq .250\ell^4$: as it was for for the cylinder, the integral is
completely dominated by ``head-on'' collisions ($98\%$ of the
integral), due to the rapid decay of the displacement with impact
parameter.  For both the cylinders and spheres, the logarithmic
singularity is the dominant contribution to the integral.  The
coefficient of the logarithm in Eqs.~\eqref{eq:Deltacylasym}
and~\eqref{eq:Deltasphasym} is given by the linearized flow near the
stagnation points at the front and rear of the cylinder or sphere,
suggesting that the integral is easy to approximate for more
complicated swimmers.  Putting all the numerical factors together, we
find the effective diffusivity
\begin{equation}
  \kappa = \begin{cases}
    1.19\, \Uc\nd\ell^3 & \text{(cylinders)};\\
    .266\, \Uc\nd\ell^4 & \text{(spheres)}.
  \end{cases}
  \label{eq:diffcylsph}
\end{equation}
We can justify these formulas dimensionally by observing that the
frequency of collisions is linear in both~$\Uc$ and~$\nd$, and since
we are assuming the path length is infinite the only other length
scale is the swimmer size~$\ell$.

\begin{figure}
  \begin{center}
    \subfigure[]{
      \includegraphics[width=.45\columnwidth]{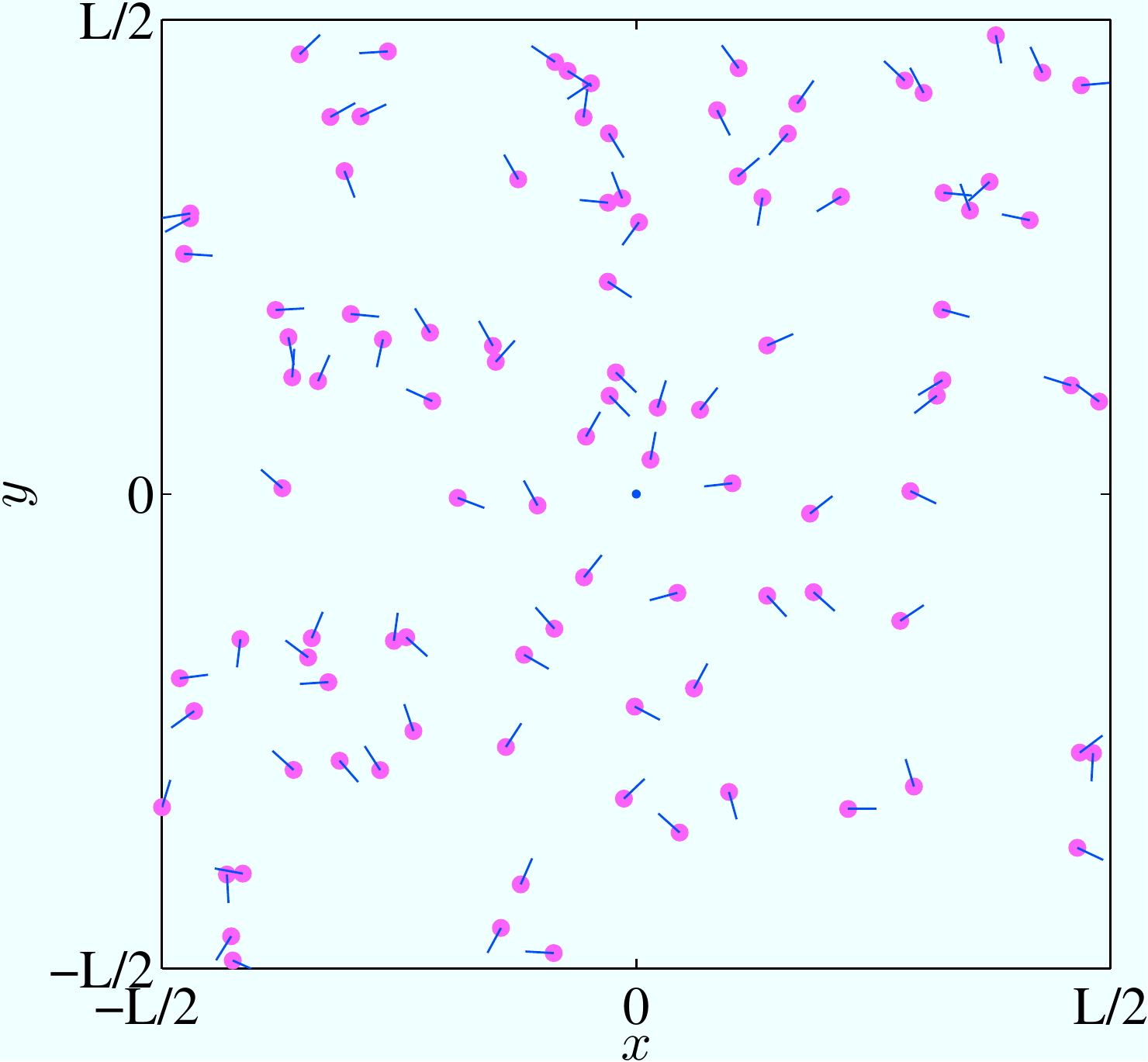}
      \label{fig:swimmers}
    }\hspace{1em}%
    \subfigure[]{
      \includegraphics[width=.44\columnwidth]{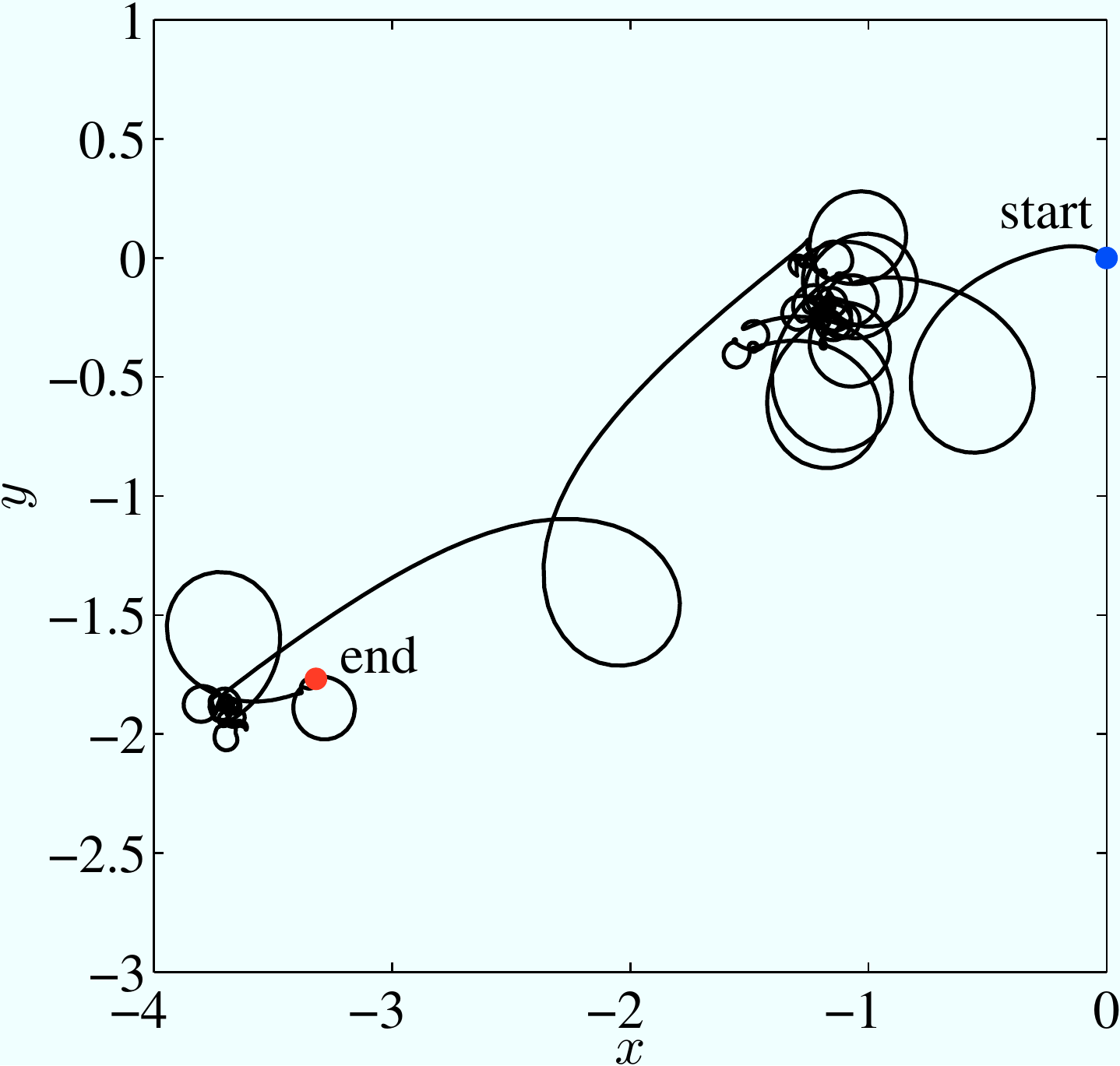}
      \label{fig:onetraj}
    }
  \end{center}
  \caption{(a) Initial configuration of~$\Nswim=100$ cylinders moving
    at constant speed~$\Uc=1$ in random directions.  Each cylinder has
    radius~$\lsc=1$ and the periodic box size used in the simulation
    is~$\Lsc=1000$ (cylinders not to scale).  (b) Trajectory of target
    particle, initially at the origin, integrated for~$10^5$ time
    units.}
\end{figure}

\textsl{Direct simulation of dilute suspensions.}  To validate
our theoretical predictions, we consider the encounters of 2D swimmers
moving in straight lines within the square \Square: $-\tfrac12\Lsc\leq
x,y\leq \tfrac12\Lsc$. We assume each swimmer is a cylinder of radius
$\lsc = 1$, with~$\Lsc\gg\lsc$. We initially place $\Nswim$ swimmers
at random positions within \Square, which subsequently move with unit
speed in a random direction.  (Figure~\ref{fig:swimmers} shows a
typical initial configuration.)  Positions are subsequently computed
mod~$\Lsc$ in both directions, maintaining the number density $\nd$ in
\Square. Diluteness requires that $\nd\lsc^2 \ll 1$.  A target
particle initially at the origin moves under the potential flow
created by all of the cylinders.  Since the cylinders are typically
well-separated, we compute their net velocity field by linear
superposition.  We show an example of this computation in
figure~\ref{fig:onetraj}, for the values $\Lsc=1000$, $\nd=10^{-4}$.
The larger ``ribbons'' caused by drift are easily identified,
suggesting that in this dilute limit the approximation of encounters
as being independent will hold.

\begin{figure}
  \begin{center}
    \includegraphics[width=.8\columnwidth]{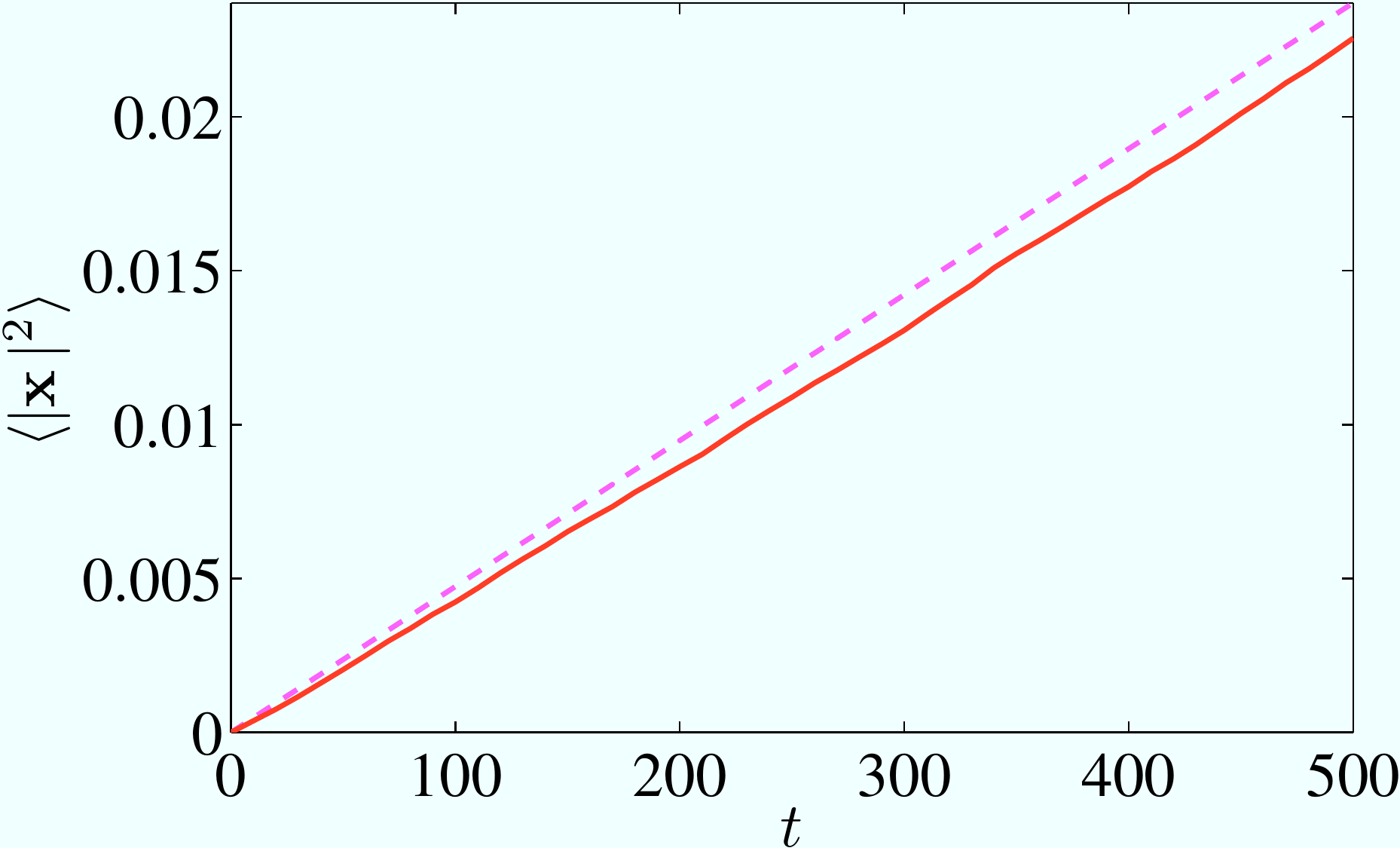}
  \end{center}
  \caption{The mean-squared displacement (solid line) of a target
    particle for $2\times 10^6$ realizations of $\Nswim=10$
    cylinders, with otherwise the same parameters as in the caption to
    Fig.~\ref{fig:swimmers}.  The dashed line shows the
    squared-displacement predicted by Eq.~\eqref{eq:squaredisp2D3D}
    and~\eqref{eq:diffcylsph}, using a number density~$\nd=10^{-5}$.
    The discrepancy between the two lines is due
    to~\eqref{eq:diffcylsph} being valid in the limit of infinite
    dilution.
  }
  \label{fig:trials}
\end{figure}

Figure~\ref{fig:trials} shows the mean-squared
displacement~$\<|\xv|^2\>$ of a target particle over $2\times
10^6$ trials (realizations) for~$\Nswim=10$ cylinders, again with unit
radius and speed.  The solid line confirms that~$\<|\xv|^2\>$ grows
linearly with time, and the dashed line shows the 2D theoretical
prediction~\eqref{eq:diffcylsph} for cylinders.  The discrepancy is
due to Eq.~\eqref{eq:diffcylsph} only being valid in the limit of
infinite dilution, i.e. as~$\nd\lsc^2\rightarrow 0$.  We have verified
that as the dilution is increased the slope approaches the theoretical
prediction.
The mean is dominated by a few trajectories with large displacements,
corresponding to small impact parameter~$\ai$.  Analogous remarks
apply to a suspension of spheres in three dimensions: the theoretical
predictions are verified there as well.


\textsl{Typical numerical values.}  We will use values for typical
krill as in~\cite{Visser2007}.  We consider spheres of
radius~$\ell=1\cm$, swimming speed~$\Uc=5\cm/\sec$, and number
density~$\nd=5\times 10^{-3}\cm^{-3}$.  Equation~\eqref{eq:diffcylsph}
then gives an effective diffusivity of~$7\times 10^{-3}\cm^2/\sec$,
about five times the thermal molecular value~$1.5 \times
10^{-3}\cm^2/\sec$, and five hundred times the molecular value~$1.6
\times 10^{-5}\cm^2/\sec$ for salt.  This implies a considerable
enhancement to the molecular diffusion, but we emphasize that these
values apply within a school of krill: the distribution and size of
the schools themselves is a more complicated
matter~\cite{Huntley2004}.  Note also that a small change in the
swimmer size has a huge impact: for a radius of~$.5\cm$, the effective
diffusivity is~$4\times 10^{-4}\cm^2/\sec$, an order of magnitude
smaller than for~$1\cm$.  If we use mean densities as discussed in the
introduction, the effective diffusivity decreases by a factor of~$10$
to~$100$.

\textsl{Effect of viscosity.}  We expect viscosity to greatly
enhance~$\kappa$.  This will be the focus of future investigation,
but for now we present a rough estimate of the impact of viscous
no-slip boundary conditions at the surface of the swimmer.  For
inviscid flow, the displacement function~$\Delta(\ai)$ has a
logarithmic singularity near the axis of swimming; for viscous flow
near a no-slip boundary, the displacement function has the stronger
singularity
\begin{equation}
  \Delta(\ai) \sim C\ell^2 / \ai,\qquad \ai \ll \ell,
  \label{eq:dispvisc}
\end{equation}
where~$C=\sqrt{2/3}\,\pi$ for a sphere of
radius~$\ell$~\cite{Young1991, Eames2003b}.  This implies that the 3D
squared-displacement integral in~\eqref{eq:squaredisp2D3D} diverges
as~\hbox{$\ai\rightarrow 0$}.  The divergence of the second
moment~$\<|\xv|^2\>$ is often associated with Levy flights, but here
we are interested in scales that are much larger than the typical
correlation length of swimming, \ie, the typical length~$\pal$ for
which a swimmer travels roughly in a straight line before changing
direction.  We thus expect the overall long-time transport to remain
diffusive, and we can cap-off the displacement function at a maximum
value~$\pal$.  In other words, a particle which is directly in the
path of a swimmer cannot travel further than the swimmer itself: this
regularizes the integral~\eqref{eq:squaredisp2D3D} to give
\begin{equation}
  \<|\xv|^2\> \simeq
    2\pi \Uc\nd\, t \int_{C\ell^2/\pal}^\infty \ai\,\Delta^2(\ai)d\ai
\end{equation}
where the lower bound of the integral is the value at which the
displacement~\eqref{eq:dispvisc} achieves its maximum allowable
value,~$\pal$.  We introduce a transition length scale where we switch
from the boundary-layer form~$\Delta\sim\ai^{-1}$ to the inviscid form
derived earlier, and find again that the dominant contribution to the
integral arises from small~$\ai$, as was the case for potential flow,
yielding
\begin{equation}
  \kappa \simeq
  \tfrac{\pi}{3}\,\Uc\nd\ell^4\,C^2 \log(\pal/\ell).
\end{equation}
For spheres, this is~$\kappa \simeq
({2\pi^3}/{9})\Uc\nd\ell^4\,\log(\pal/\ell) =
6.89\,\Uc\nd\ell^4\,\log(\pal/\ell)$.  Inserting the same numerical
values for krill as before, with a path length~$\pal=100\cm$, we
find~$\kappa \simeq .8 \cm^2/\sec$, about~$500$ times the molecular
value.  Thus, including the effect of viscosity and finite path length
has increased the effective diffusivity by a factor of~$100$ over the
inviscid flow case.  We emphasize that this is a rough estimate.  The
path length (or swimming correlation length) is a measure of how much
a swimmer tends to move in one direction before turning.  Our chosen
value of~$1\meter$ is not based on any evidence, but~$\kappa$ has only
a weak logarithmic dependence on~$\pal$.  Assuming a
value~$\pal=10\meter$ only raises~$\kappa$ from~$.8$
to~$1.2\cm^2/\sec$.

Any conclusion regarding the importance of biomixing in the oceans
must be carefully qualified: at the densities inside of schools, the
inclusion of viscous effects suggests a rather large enhanced
diffusivity, comparable with other processes~\cite{Munk1966}, while
outside of schools the effect is much weaker.  However, our viscous
estimate is rough and more effects will need to be included to form a
complete theory: boundary layers, more realistic shape distributions
for the swimming bodies, wakes, spatial correlations between the
swimmers, patchiness and schooling, finite correlation length of
swimming, distribution of velocities, and buoyancy and stratification
effects.  This last item is probably the most important:
stratification can cause fluid parcels to return to their initial
height after being displaced if they can't equilibrate their density
with their surroundings.  A mechanism such as enhanced diffusion due
to small-scale turbulence might assist this equilibration.

The simplicity of our model means that prefactors and scalings can be
computed accurately.  The numerical constants we obtained depend
mostly on the flow near the stagnation points around the swimming
body.  Our simple model can serve as a platform on which to build
complexity, or could be applied to other fluid-dynamical systems where
a collection of objects causes mixing, such as in sedimentation.

\begin{acknowledgments}
  The authors are grateful to W. Dewar, R. Ferrari, M. Graham,
  Z.~G. Lin, C. Ortiz--Duenas, Y.-K. Tsang, and W. Young for helpful
  discussions, as well as to the hospitality of the 2008 Summer
  Program in Geophysical Fluid Dynamics (supported by NSF and ONR) at
  WHOI, where this work began, and the Institute for Mathematics and
  its Applications (supported by NSF).  SC was supported by NSF under
  grant DMS-0507615, J-LT under grant DMS-0806821.
\end{acknowledgments}


\end{document}